\documentclass[prd,amsmath,amssymb,preprintnumbers,twocolumn,nofootinbib,10pt]{revtex4-1}
\pdfoutput=1
\usepackage{graphicx}
\usepackage{dcolumn}
\usepackage{bm}
\usepackage{amssymb}
\usepackage{latexsym}
\usepackage{booktabs}
\usepackage{amsmath}
\usepackage{multirow}
\usepackage{url}
\usepackage{footnote}
\usepackage{float}
\usepackage{threeparttable}
\usepackage[colorlinks=true, linkcolor=blue, citecolor=blue]{hyperref}
\usepackage[bottom]{footmisc}

\usepackage[normalem]{ulem}
\usepackage{color}
\usepackage{array}
\usepackage{enumerate}
\usepackage{adjustbox}

\usepackage{makecell}
\usepackage{diagbox}
\usepackage{epstopdf}
\usepackage{epsfig}
\usepackage{longtable}
\usepackage{supertabular}
\usepackage{algorithm}
\usepackage{pifont}
\usepackage{algorithmic}
\usepackage{changepage}
\usepackage{setspace}

\begin{document}

\title{CMB delensing with deep learning}

\author{Shulei Ni$^{1,2}$, Yichao Li$^{1}$ and Xin Zhang$^{1,3,4}$}
\email{zhangxin@mail.neu.edu.cn}

\affiliation{$^1$Liaoning Key Laboratory of Cosmology and Astrophysics, College of Sciences, Northeastern University, Shenyang 110819, China}
\affiliation{$^2$Research Center for Astronomical Computing, Zhejiang Laboratory, Hangzhou 311121, China}
\affiliation{$^3$MOE Key Laboratory of Data Analytics and Optimization for Smart Industry, Northeastern University, Shenyang 110819, China}
\affiliation{$^4$National Frontiers Science Center for Industrial Intelligence and Systems Optimization, Northeastern University, Shenyang 110819, China}


\begin{abstract}
The CMB stands as a pivotal source for studying weak gravitational lensing. While the lensed CMB aids in constraining cosmological parameters, it simultaneously smooths the original CMB's features. The angular power spectrum of the unlensed CMB showcases sharper acoustic peaks and more pronounced damping tails, enhancing the precision of inferring cosmological parameters that influence these aspects. Although delensing diminishes the {\rm B}-mode power spectrum, it facilitates the pursuit of primordial gravitational waves and enables a lower variance reconstruction of lensing and additional sources of secondary CMB anisotropies. We employed the UNet++ algorithm to perform operations and analysis on CMB delensing, presenting the angular power spectra of {\rm TT}, {\rm EE}, and {\rm BB} after CMB delensing, and compared them with those obtained using the QE delensing algorithm.
The lensing CMB sky map and full-sky angular power spectrum processed by the UNet++ algorithm are very close to those of the CMB without lensing effects, and the error is more than 10 times smaller than that given by the QE algorithm.
The code utilized for this analysis is publicly available.
\end{abstract}

\maketitle

	\section{Introduction} \label{sec:intro}

The anisotropies of the cosmic microwave background (CMB) offer crucial insights
into the early universe. Observations of such anisotropies have made significant contributions to 
establishing the current standard model of cosmology.
In the last few decades, three generations of satellite experiments, namely 
{\it COBE}~\cite{COBE:1992syq, Fixsen:1996nj,Bennett:1996ce},
{\it WMAP}~\cite{WMAP:2003elm,WMAP:2008lyn,WMAP:2010qai}, and
{\it Planck}~\cite{Planck:2013pxb,Planck:2015fie,Planck:2018vyg}, 
as well as numerous ground-based experiments
(e.g., {DASI}~\cite{Leitch:2002fe}, {CBI}~\cite{Readhead:2004xg}, {SPTpol}~\cite{Carlstrom:2009um},
{BICEP}~\cite{BICEP1:2013sbv,BICEP2:2014dgt}, {Keck Array}~\cite{Staniszewski:2012eqo}, 
{ACT}~\cite{ACT:2023wcq}, and {Ali-CPT}~\cite{Li:2017drr}) and balloon experiments 
(e.g., {BOOMERanG}~\cite{Masi:2002hp}, {EBEX}~\cite{EBEX:2017etk}, {SPIDER}~\cite{Crill:2008rd}), 
have been carried out to precisely measure the temperature and polarization power spectrum of the CMB.
Moreover, there are several ongoing plans for diverse experimental projects, 
including {\it LiteBIRD}~\cite{LiteBIRD:2022cnt}, {CMB-S4}~\cite{CMB-S4:2016ple, CMB-S4:2022ght}, {AliCPT}~\cite{Li:2017drr},
and {PIPER}~\cite{Lazear:2014bga}, being devoted to the detection of CMB polarization signals.

The generation of the stochastic background of gravitational waves, known as 
primordial gravitational waves (PGWs), is a fundamental prediction in any cosmological inflation model~\cite{Lyth:1998xn,Baumann:2009ds}. The characteristic of this signal encodes unique 
information about the physics of the early universe and its subsequent evolution, 
providing an exciting and powerful window into the universe's origin and evolution.

The tensor-to-scalar ratio $r$ parameterizes the amplitude of PGWs and connects to the energy scale at which inflation occurred. Therefore, the detection of PGWs is expected to reveal the physical properties of the universe's early stages. Fortunately, these PGWs have distinct imprints in the polarized anisotropy of the CMB, displaying a spiral-like {\rm B}-mode pattern that enables the extraction of the PGW signal from the polarized {\rm B}-mode of the CMB~\cite{BICEP2:2014owc,CMB-S4:2016ple}. The detection of {\rm B}-mode signals presents an exceptionally challenging task, particularly due to the uncertain extent of foreground contamination and the mixing of relatively strong {\rm E}-mode signals with {\rm B}-mode signals induced by weak gravitational lensing~\cite{BICEP2:2014owc}. Therefore, the primary scientific objectives of CMB polarization observations currently entail determining the extent of foreground contamination and precisely measuring the effect of gravitational lensing~\cite{SimonsObservatory:2018koc}. Our research primarily centers on the gravitational lensing phenomenon associated with CMB.

CMB lensing has been extensively studied in theory~\cite{Lewis:2006fu,Okamoto:2003zw,Bartelmann:1999yn}. After decoupling from matter at the last scattering surface, CMB photons propagate freely and are gravitationally deflected by the universe's large scale distribution of matter. The lensing effect results in subtle imprints on the CMB temperature and polarization anisotropies. These can be utilized to create a map of the lensing potential that determines the lensing deflections' gradient~\cite{Hu:2000ee}. Weak gravitational lensing smoothes the acoustic peaks of the CMB angular power spectra, transferring power from larger angular scales to smaller scales, and converting {\rm E}-mode polarization to {\rm B}-mode polarization \cite{Zaldarriaga:1998ar,DODELSON2021373}.
In addition, weak gravitational lensing can induce small distortions in the CMB, and these distortions can be detected by modifying its primordial morphology in an anisotropic form. 

Given that the anisotropy during the final scattering can be approximated as Gaussian, the non-Gaussian structures observed in the lensed CMB sky provide additional information about the properties of gravitational lensing material. This is crucial for a further understanding of important details regarding the distribution of matter~\cite{Lewis:1999bs, Lewis:2006fu, Lewis:2013hha}. Therefore, weak gravitational lensing of the CMB both promotes and hinders our comprehension of the history and content of the universe.

There are several advantages associated with mitigating the impacts of lensing on the observed CMB temperature and polarization maps~\cite{Lewis:2005tp, Lewis:2006fu,Lewis:2013hha,DODELSON2021373}. Firstly, the average temperature of the CMB remains unchanged by lensing as the gravitational effect only adjusts arrival directions and not the surface brightness~\cite{Lewis:2005tp,Lewis:2013hha}. Secondly, the distortion of light paths traveling from far sources to reach us is caused by the gravitational effect of universe's inhomogeneities. The reason lensing is so promising is that it enables probing all clustering stress-energy components in the universe through spacetime perturbations since light paths react to mass. Measurement of these distortions provides insight into the mass distribution of the universe~\cite{Lewis:2005tp, Lewis:2006fu, Lewis:2013hha}. 

However, gravitational lensing has the potential to smoothen acoustic peaks in the CMB. This can be comprehended by considering that when photons get deflected, features of a fixed angular size can either be magnified or de-magnified, causing sharp features in the power spectrum to be blurred across a range of scales. The angular scale of sharp features in the power spectrum is easier to measure than broad humps; consequently, gravitational lensing weakens our ability to precisely measure acoustic peak positions in the CMB power spectra~\cite{Lewis:2005tp,Lewis:2006fu}. The temperature power spectrum undergoes lensing by $0.2\%$ at $\ell \sim 2000$, but smaller scales experience changes at the percentage level. The impact on the {\rm B}-mode polarization power spectrum is more significant, with power increase of $6\%$ across all scales~\cite{Lewis:2006fu}.

Delensing reverses this peak smoothing, providing sharper peaks with a more precisely measurable angular scale. Similar observations can be made about the measurement of peak heights. Weak gravitational lensing alters the CMB power spectra, induces non-Gaussian, and generates a {\rm B}-mode polarization signal, which causes confusion for the signal from PGWs~\cite{Lewis:2005tp,Lewis:2006fu}.

In recent years, deep learning and artificial intelligence techniques have developed rapidly, gaining widespread attention and significant application in various disciplines. Astronomy research has been catching up, and many studies have emerged, applying deep learning for data analysis~\cite{Gupta:2018eev,Caldeira:2018ojb, zhou2018unet++,Zhang:2019ryt,zhou2019unet++,Springer:2018aak,Makinen:2020gvh,Guzman:2021ygf,Ni:2022kxn,Gao:2022xdb,Yan:2024czw}. 
These studies have demonstrated the high effectiveness of deep learning in image reconstruction and segmentation tasks, enabling the detection of features at the pixel level~\cite{wang2018interactive,ghosh2019understanding,minaee2021image}. 

This work adopts the sky map segmentation method proposed by Makinen~\cite{Makinen:2020gvh}. The method relies on the HEALPix pixelization scheme, which divides the two-dimensional spherical sky map into multiple two-dimensional plane images. We apply this sky map segmentation method to attempt a novel CMB delensing approach that can remove the lensing effect.

The paper is organized as follows. In Section \ref{sec:simulation}, we present the simulation of CMB and the preprocessing of simulated data. In Section \ref{sec:unet}, we describe the methods of quadratic estimator~(QE) delensing and UNet++ delensing. In Section \ref{sec:result}, we provide a comprehensive analysis of the obtained results. Finally, Section \ref{sec:conclusions} provides the concluding remarks.

%%%%%%%%%%%%%%%%%%%%%%%%%%%%%%%%%%%%%%%%%%
\section{Data Simulation and Pre-processing}\label{sec:simulation}

In this section, we will describe the simulated CMB and noisy sky maps used for our analysis. This includes detailed information on lensed temperature and polarization maps with noise and instrumental effects, as well as unlensed temperature and polarization maps.

For our simulations, we employed the publicly available package, \href{https://github.com/carronj/lenspyx}{Lenspyx}~\cite{Lewis:2005tp, Diego-Palazuelos:2020lme}, a Python package specifically designed for simulating lensed CMB maps on a curved sky. We also relied on three additional software packages, namely \href{https://camb.info/}{CAMB} \cite{Lewis:1999bs}, \href{https://cosmologist.info/lenspix/}{LensPix}~\cite{Lewis:2005tp} and \href{https://github.com/healpy/healpy/}{Healpy} \cite{Gorski:2004by, Zonca:2019vzt}, as the foundational components for developing detailed curved-full-sky simulations of both lensed and unlensed CMB. By incorporating the {\it Planck} 2018 CMB lensing pipeline (plancklens), Lenspyx has the ability to replicate both the published map and band-powers. 

We have employed a concise formula to generally describe our data simulation process, which is presented as follows:
\begin{equation}\label{eq:sim_data}
	\mathcal{M}^\mathcal{F}_{{\rm lensed}}= {\rm Smooth}[{\rm Lenspyx}(\mathcal{M}^\mathcal{F}_{{\rm unlensed}}) + \mathcal{M}^\mathcal{F}_{\rm N}],
\end{equation}
where, $\mathcal{M}$ represents the full sky map, and $\mathcal{F}\in\{{\rm T}, {\rm Q}, {\rm U}\}$. ${\rm Smooth}$ represents the telescope's beam, and we have chosen a simple Gaussian beam. Lenspyx is an abbreviation for the process of converting unlensed maps into lensed maps. $N$ represents the noise component in the image. We also considered the impact of instrumental effects and used a Gaussian beam with a full width at half maximum~({\rm FWHM}) of $7.9$ {\rm arcmins} to simplify the processing.

\subsection{The CMB temperature and lensing potential}\label{subsec:TT}
The CMB radiation field is represented by the temperature anisotropy, denoted as $T(\hat{n})$, and the polarization, denoted as $P(\hat{n})$, in the spatial direction $\hat{n}$ of the celestial sphere. We observe CMB temperature changes projected on a 2D spherical surface sky, and it is now habitual in the literature to expand the temperature field using spherical harmonics. 

The temperature fluctuation of the CMB  on the spherical surface is described by a scalar field, consisting of small fluctuations $\Delta T(\hat{n})$ at a level of $10^{-5}$ relative to the average value $T_0 = 2.725$ K. For full-sky observations, the temperature field {\rm T} can be expressed through the spherical harmonic decomposition using spin-0 spherical harmonics $Y_{lm}(\hat{n})$,
\begin{equation}\label{T_temp}
	\Delta T(\hat{n}) = \sum_{\ell=0}^{l_{max}}\sum_{m=-\ell}^{m=\ell}a_{\ell m}Y_{\ell m}(\hat{n}).
\end{equation}
All the information contained in the temperature field $T(\hat{n})$ is included in the space-time dependent amplitudes $a_{\ell m}$, where $a_{\ell m}$ represents the spherical harmonic coefficient and can be expressed using the following formula,
\begin{equation}
	a_{\ell m}=\int {\rm d}\Omega_n\Delta T(\hat{n})Y_{\ell m}^*(\hat{n}),
\end{equation}
and analogously to the methodology in Fourier space, we can define an angular power spectrum for these fluctuations, denoted as $C_\ell$, by calculating the variance of the harmonic coefficients,
\begin{equation}\label{eq:t_alm}
	<a_{\ell m}a_{\ell'm'}^*>=\delta_{\ell\ell'}\delta_{mm'}C_\ell,
\end{equation}
where the above average is taken over many ensembles and the delta functions arise from isotropy. We can write the following expression for the angular power spectrum,
\begin{equation}
	C_\ell^{TT}=\frac{1}{2\ell+1}\sum_{m}<a_{\ell m}^T{a_{\ell m}^T}^*>.
\end{equation}

We can simulate the angular power spectrum $C_\ell$ with CAMB or \href{https://lesgourg.github.io/class\_public/class.html}{CLASS}~\cite{Lucca:2019rxf, DiDio:2013bqa}, as shown in Fig.~\ref{fig:BB_aps}. The figure demonstrates that as the scale decreases, the difference between lensed and unlensed data amplifies. This phenomenon occurs due to the deflection imparted on CMB photons by each encountered potential, resulting in an accumulation of effects on the CMB power spectrum at small scales. Consequently, the lensing effect distorts the original power spectrum and introduces non-Gaussianity in the lensed CMB~\cite{Lewis:2006fu}. It should be noted that non-linear evolution also contributes to an increase in power on smaller scales.

Weak lensing of the CMB deflects photons coming from an original direction $\hat{n}'$ on the last scattering surface to direction $\hat{n}$ on the observed sky, so a lensed CMB temperature field, $T(\theta, \phi)$, is given by $\widetilde{X}(\hat{n})=X(\hat{n}')$ in terms of the unlensed field $X=T$~\cite{Lewis:2005tp}. Thus the position in the sky where we finally see the CMB photons is determined by the integral of the gravitational potential along the line of sight to the last scattering surface.

We propose the definition of an integrated lensing potential, denoted as $\psi$. The deflection vector is expressed as the gradient of the lensing potential, ${\rm \nabla}\psi(\hat{n})$, where ${\rm \nabla}$ represents the covariant derivative on the sphere. The vector $\hat{n}'$ is derived from $\hat{n}$ by moving its geodesic at one end of the surface of the unit sphere along the $\hat{n}'$ direction by a distance ${\rm \nabla}\psi(\hat{n})$. Then the unlensed CMB photon with direction $\hat{n}$ becomes a lensed CMB photon with direction $\hat{n}=\hat{n}' + {\rm \nabla}\psi(\hat{n})$, through the weak gravitational lensing. Thus, the lensed CMB temperature can be written as follows,
\begin{equation}
	\widetilde{T}(\hat{n})=T(\hat{n}')=T(\hat{n}+{{\rm \nabla}}\psi(\hat{n})),
\end{equation}
where lensing potential $\psi(\hat{n})$ can be defined as~\cite{Lewis:2006fu, Hassani:2015zat},
\begin{equation}
	\psi(\hat{n})=2\int_0^{\chi_\star} {\rm d}\chi(\frac{\chi_\star-\chi}{\chi_\star\chi})\Psi(\hat{n}, \eta),
\end{equation}
where $\eta$ is the conformal time, $\chi$ is the comoving distance and $\Psi$ is the Bardeen potential. As with the CMB temperature angular power spectrum, the same angular power spectrum of the gravitational potential can be obtained,
\begin{equation}\label{equ:phi_cl}
	C_\ell^{\psi\psi}=16\pi\int\frac{{\rm d}k}{k}P_R(k)(\int_0^{\chi_\star}{\rm d}\chi(\frac{\chi_\star-\chi}{\chi_\star\chi})T(k, \eta)k_\ell(k\chi))^2,
\end{equation}
where $T(k, \eta)$ is the appropriate transfer function, $P_R(k)$ is the primordial power spectrum.

We utilized Lenspyx to simulate lensed CMB and unlensed CMB temperature and polarization sky maps. In our simulation, we utilized the best-fit cosmological parameters from the {\it Planck} 2018 $\Lambda$CDM results~\cite{Planck:2018vyg}, i.e., $H_0=67.7$ km s$^{-1}$ Mpc$^{-1}$, $\Omega_{\rm b}=0.049$, $\Omega_{\rm m}=0.311$, $\Omega_\Lambda=0.689$, and $\sigma_8=0.81$. For the simulation of the temperature sky map, we ran the parameters $A_s$ and $n_s$, as they are more sensitive to it, while for the polarization, we only concentrate on the tensor-to-scalar ratio. Both temperature and polarization simulations were conducted autonomously.

To acquire additional datasets and prepare for subsequent polarization analysis, we conducted experiments using two key cosmological parameters: the scalar amplitude ($A_s$) and the tensor-to-scalar ratio ($r$), both of which play crucial roles in determining the temperature and polarization power spectra of the CMB. Specifically, we sampled five values of $A_s$ ($2.0 \times 10^{-9}$ to $2.2 \times 10^{-9}$) and ten values of $r$ linearly spaced between $0.001$ and $0.01$, resulting in a total of 30 CMB TT angular power spectra and corresponding sky maps.

In the top panel of Figure~\ref{fig:BB_aps}, the red solid line and the pink dashed line illustrate the lensed {\rm TT} angular power spectrum and the unlensed {\rm TT} power spectrum , with the parameters $A_s = 2.1 \times 10^{-9}$ and $r = 0.005$, respectively. The ratio between them is shown with a red dashed line in the bottom panel. On large scales, the anisotropies of the CMB are primarily dominated by emissions from the last scattering surface at a redshift of $z\sim1100$. However, on smaller scales, the CMB is more significantly influenced by what are known as secondary effects. These secondary anisotropies arise as a result of interactions between CMB photons and matter along the line of sight.

Our work is grounded in the analysis of full-sky maps, thus requiring the use of the Lenspyx tool to transform the generated $30$ angular power spectra into corresponding temperature sky maps. To investigate the impact of lensing potential on small-scale structures, we conducted simulation experiments at a high resolution of $N_{\rm side}=2048$. To ensure that the simulation results more closely match actual observational data, we accounted for the smoothing effect of the telescope beam on signals during the computation process, use a {\rm FWHM} of $\theta_{{\rm FWHM}}=8.3$ {\rm arcmins}. 

\subsection{The CMB polarization}\label{subsec:BB}
CMB polarization is measured through time-averaged Stokes parameters -- measures of linear polarization of the electric field aligned orthogonally to the line of sight's (LOS) Cartesian axes. Due to the nature of Thomson scattering, {\rm Q} and {\rm U} are ample in describing CMB polarization, given their inability to generate circular polarization~\cite{DODELSON2021373}.
Astronomical observations reveal that Stoke parameters {\rm Q} and {\rm U} are linked by a constant and relative $45{\rm ^\circ}$ degree rotation around LOS -- whereas the reference frame may revolve freely around it.
Bundling the reference frame with a fixed set of normal vectors in rotation will directly link {\rm Q} and {\rm U} to the {\rm E} and {\rm B} modes.

In this subsection, we provide a concise overview of the polarization field properties and its decomposition into physically distinct {\rm E} and {\rm B} modes. We express the standard construction of {\rm E} and {\rm B} fields in terms of the spin-raising and spin-lowering operators, usually implemented in harmonic space.

If we rotate {\rm Q} and {\rm U} by an angle $\alpha$ on the plane that is perpendicular to the direction of $\hat{n}$, we obtain the following solution~\cite{Zaldarriaga:1996xe, zaldarriaga1998cosmic,Rotti:2018pzi},
\begin{equation}
	(Q\pm iU)'(\hat{n})=e^{\mp2i\alpha}(Q\pm iU)(\hat{n}).
\end{equation}
We can derive the separation of {\rm E} and {\rm B} modes from the Stokes parameters~\cite{Kamionkowski:1996ks, Kamionkowski:1996zd, Bunn:2002df, Kamionkowski:2015yta, kim2010b}. Here we provide a brief overview of the standard method. Additionally, we can decompose {\rm Q} and {\rm U} into $\pm 2$ spin spherical harmonics concerning rotation as shown below~\cite{Zaldarriaga:1996xe, zaldarriaga1998cosmic},
\begin{equation}
	Q(\hat{n})\pm iU(\hat{n})=\sum_{\ell,m} a_{\pm 2,\ell m}{_\pm}{_2}Y_{\ell m}(\hat{n}),
\end{equation}
where ${_\pm}{_2}Y_{\ell m}(\hat{n})$ are the spin $\pm 2$ spherical harmonics,  and the coefficients $a_{\pm 2, \ell m}$ are given by
\begin{equation}
	a_{\pm2, \ell m}=\int(Q(\hat{n})\pm iU(\hat{n})){_\pm}{_2}Y_{\ell m}^\star(\hat{n}){\rm d}\hat{n}.
\end{equation}
The {\rm E} and {\rm B} modes in the spherical harmonic space are formed by
\begin{equation}\label{eq:eb_alm}
	\begin{split}
		&a_{\ell m}^E=-(a_{2,\ell m}+a_{-2,\ell m})/2,\\
		&a_{\ell m}^B=-(a_{2,\ell m}-a_{-2,\ell m})/2i.
	\end{split}
\end{equation}
By applying the angular correlation function, the sum can be reduced to an expression that only involves $\ell$ and power spectrum term
\begin{equation}
	\begin{split}
		&C_\ell^{TE}=\frac{1}{2\ell+1}\sum_m<a_{\ell m}^T{a_{\ell m}^E}^*>,\\
		&C_\ell^{EE}=\frac{1}{2\ell+1}\sum_m<a_{\ell m}^E{a_{\ell m}^E}^*>,\\
		&C_\ell^{BB}=\frac{1}{2\ell+1}\sum_m<a_{\ell m}^B{a_{\ell m}^B}^*>.
	\end{split}\label{EB_pol}
\end{equation}

Based on the aforementioned theoretical derivations, we employed the same methodology used for simulating temperature sky maps to simulate polarized sky maps. We conducted tests on $30$ datasets utilizing two critical parameters, the equation-of-state parameter of dark energy $w$ ($-1.025$, $-1$, $-0.975$) and tensor-to-scalar ratio $r$ ($0.001$, $0.002$, $0.003$, $0.004$, $0.005$, $0.006$, $0.007$, $0.008$, $0.009$, $0.01$). Similar to the simulation of temperature sky maps, each distinct set of parameters will be employed to generate a corresponding full-sky map.

\begin{figure}
	\centering
	\includegraphics[width=0.45\textwidth]{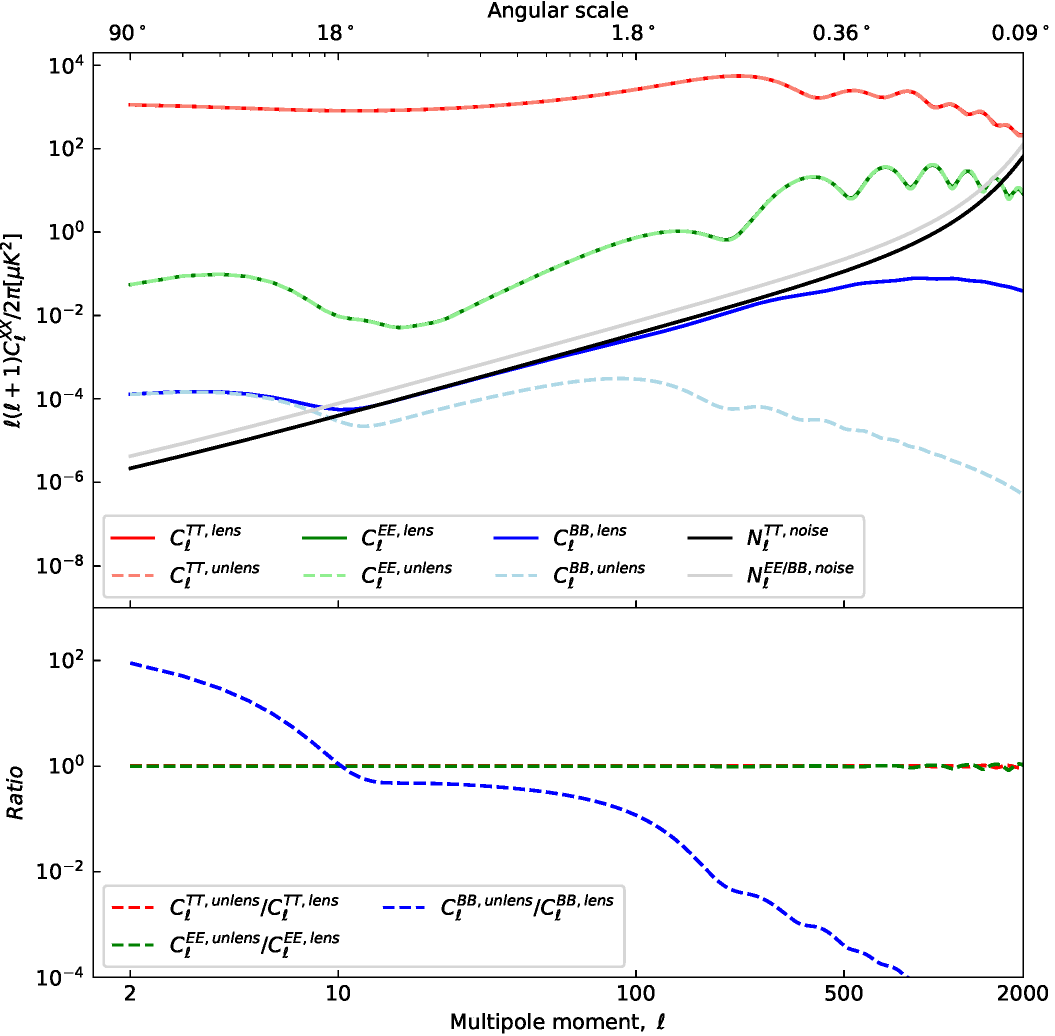}
	\caption{CMB {\rm XX} angular power spectrum, $X\in\{{\rm T}, {\rm E}, {\rm B}\}$. {\rm Top Panel:} The red solid line represents the lensed {\rm TT} spectrum, while the light red dashed line corresponds to the unlensed {\rm TT} spectrum. The green solid line denotes the {\rm EE} spectrum under the influence of lensing, whereas the light green dashed line shows the {\rm EE} spectrum without lensing effects. The blue solid line indicates the lensing-modified {\rm BB} spectrum, and the light blue dashed line depicts the {\rm BB} spectrum in the absence of lensing effects. The parameters for generating the temperature angular power spectrum are $A_s = 2.1 \times 10^{-9}$ and $r = 0.005$. The black solid line displays the noise level of the temperature detectors, and the grey solid line represents that of the polarization detectors.
		{\rm Bottom Panel:} It illustrates the relative differences between these highly similar spectra. The red dashed line marks the ratio of the unlensed {\rm TT} spectrum to the lensed {\rm TT} spectrum; the green dashed line indicates the ratio of the unlensed {\rm EE} spectrum to the lensed {\rm EE} spectrum; and the blue dashed line shows the ratio of the unlensed {\rm BB} spectrum to the lensed {\rm BB} spectrum.}
	\label{fig:BB_aps}
\end{figure}

In the top panel of Figure~\ref{fig:BB_aps}, we use a green solid line and a light green dashed line to show the lensed and unlensed angular power spectra of the E-mode, respectively, and a blue solid line and a light blue dashed line to show the lensed and unlensed angular power spectra of the B-mode, with the equation of state parameter for dark energy $w = -1$ and the scalar-to-tensor ratio $r = 0.005$, respectively. Meanwhile, in the bottom panel, a green dashed line shows the ratio of the lensed to the unlensed E-mode power spectra, and a blue dashed line shows the ratio of the lensed to the unlensed B-mode power spectra. Similar to the simulation of the temperature sky maps, we utilized the Lenspyx tool to generate $30$ sky maps corresponding to the polarization power spectra.

After obtaining the full-sky maps of the Stokes parameters {\rm T}, {\rm Q}, and {\rm U}, we transformed them into the {\rm T}, {\rm E}, and {\rm B} components in harmonic space using Equation~(\ref{eq:eb_alm}). This conversion enables a more direct analysis of the physical polarization modes and separates the scalar (E-mode) and tensor (B-mode) contributions. The resulting {\rm T/E/B} maps serve as the input for our neural network, which is specifically designed to operate on the {\rm T/E/B} fields for subsequent data processing and training.

\subsection{Noise simulation}
%This subsection presents a concise overview of noise simulation.

In the course of the actual observation of the CMB, it is crucial to consider various sources of noise interference. Given the faintness of the CMB signal, noise has the potential to mask or distort its true characteristics, leading to deviations in the measurement of core observables such as temperature and polarization. Moreover, the presence of noise can introduce systematic errors that impact our ability to infer cosmological parameters.

The noise in CMB detectors is often approximated as Gaussian white noise, and its angular power spectral density can be defined in the following~\cite{Wu:2014hta,BICEP2:2015nss,wu2020detecting, Wolz:2023lzb}:
\begin{equation}\label{eq:noise}
	{\rm N}_\ell^{XX'}=[s^2e^{(-\ell(\ell+1)\frac{\theta_{{\rm FWMH}}^2}{8\ln2})}]^{-1},
\end{equation}
where $s$ denotes the telescope's sensitivity, $X\in\{{\rm T},{\rm E},{\rm B}\}$, $\theta_{{\rm FWHM}}$ represents the {\rm FWHM} of the telescope. For simplicity, we adopt $\theta_{{\rm FWHM}} = 7.9 ~{\rm arcmin}$, with a sensitivity for the temperature detector of $1.5~{\rm \mu K}$ {\rm arcmins} and for the polarization detector of $2.1~{\rm \mu K}$ {\rm arcmins}~\cite{NASAPICO:2019thw}.

According to Equation~(\ref{eq:noise}), we can calculate the temperature and polarization angular power spectra of noise, as illustrated by the black and grey lines in the top panel of Figure~\ref{fig:BB_aps}. For the noise of each component, we generate corresponding full-sky maps based on their angular power spectra and overlay these noise maps onto the sky maps with lensing effects. Similar to the simulation of the temperature sky maps, we utilized the Lenspyx tool to generate $30$ sky maps corresponding to the noise power spectra.

\subsection{Data pre-processing}\label{data_pro}

By conducting simulations with Lenspyx, we generated 30 sets of full-sky maps for {\rm I}, {\rm Q}, {\rm U}, along with noise maps. Using Formula~\ref{eq:sim_data}, we were then able to obtain the sky maps corresponding to our training data, i.e. {\rm T}, {\rm E}, {\rm B} fields. Spherical data poses challenges for deep learning due to its non-Euclidean structure. Traditional convolutions are not directly applicable on the sphere because of the lack of translational equivariance, and the rotational symmetry on the sphere is also difficult to handle. Moreover, spherical sampling is often irregular, such as in the HEALPix grid, which complicates feature extraction.

\begin{figure}
	\centering
	\includegraphics[width=0.45\textwidth]{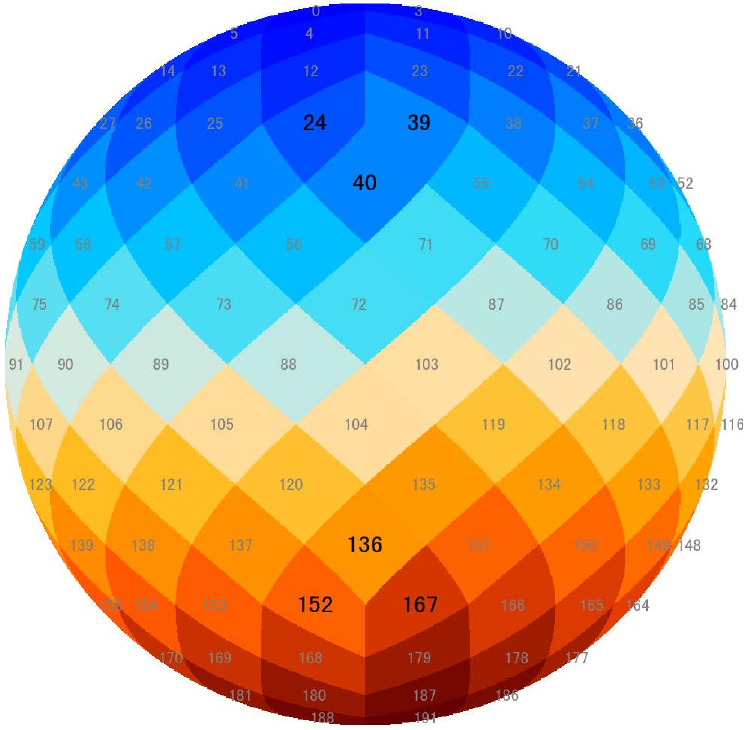}
	\caption{A half-orthogonal view projection of the sky map.Gray sequentially labels pixels with four neighbors, while larger black numbers highlight those with only three.}
	\label{fig:hp_index}
\end{figure}

\begin{figure*}
	\centering
	\includegraphics[width=0.8\textwidth]{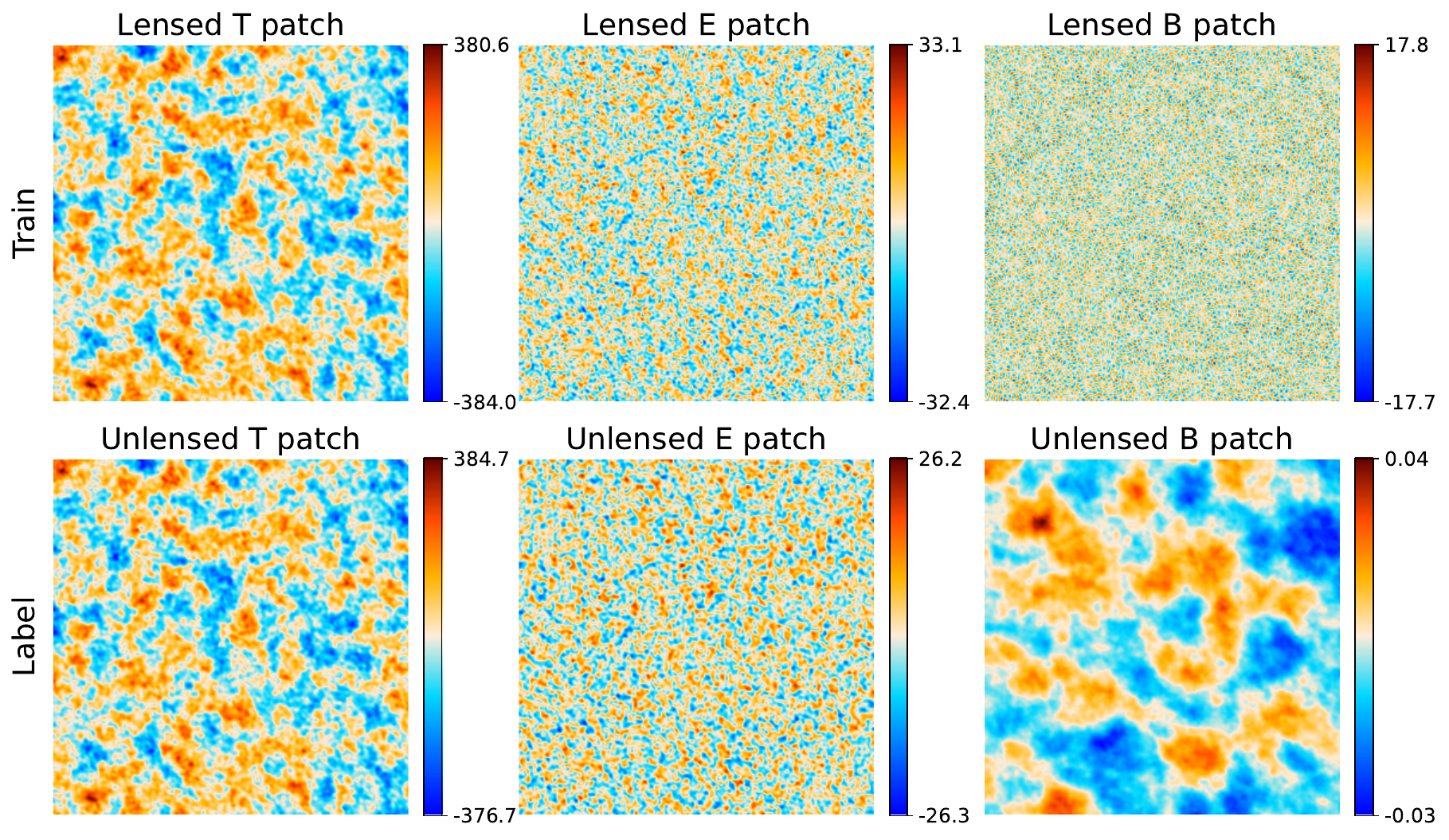}
	\caption{The {\rm T}, {\rm E} and {\rm B} sky map patch. Each patch has a size of $214.86~{{\rm deg}}^2$.
		The center of the sky patches is 
		$(l, b) = (101.25{\rm ^\circ}, 19.471{\rm ^\circ})$, where $l$ and $b$ are the
		Galactic longitude and Galactic latitude, respectively.  From left to right, the sky patches are {\rm T}, {\rm E} and {\rm B} map, respectivly. The top row of patches displays sky map patches affected by lensing effects, including noise and instrumental artifacts, which are used to construct the training dataset. The second row presents the original, unlensed sky map patches, which serve as the basis for generating the label dataset. The unit is ${\rm \mu K}$.}
	\label{fig:pathcs}
\end{figure*}

There are currently available network solutions for deep learning on the sphere, such as \href{https://github.com/deepsphere/deepsphere-cosmo-tf1}{{DeepSphere}}~\cite{Perraudin:2018rbt} and \href{https://github.com/ai4cmb/NNhealpix}{{NNhealpix}}~\cite{Krachmalnicoff:2019zjh}. Both methods are highly innovative and have made outstanding contributions to the application of artificial intelligence algorithms on spherical data. We have attempted both methods, but the extremely high resolution of the full sky map we are considering, i.e., $N_{\rm side}=2048$, makes both methods extremely time-consuming during the data initialization stage. However, regardless of the segmentation scheme used, if the full-sky map is not directly used as training data, boundary processing issues will inevitably arise. Therefore, we require more effective data processing and model training methods to address this challenge. Additionally, due to the fact that not all pixels on HEALPix have exactly four neighboring pixels, but in some cases, there are only three, this may also introduce errors in certain segmentation approaches~\cite{Krachmalnicoff:2019zjh}, as shown in Figure~\ref{fig:hp_index}.

When there is a mismatch between data and model, there are usually two ways to deal with it: modifying the model to fit the data, or adjusting the data format to meet the model requirements. In related research such as DeepSphere and NNhealpix, the first strategy is used, which is to modify the structure of the deep learning model to adapt to spherical data, so that the network can complete training. However, there are certain limitations in such methods, so we propose an alternative approach that segments spherical data to more flexibly address the mismatch between data and models. Our segmentation scheme builds upon Makinen's work~\cite{Makinen:2020gvh}. Specifically, we approximate each pixel at a resolution of $N_{\rm side}=4$ as a flat region, as illustrated by the labeled sky patches in Figure~\ref{fig:hp_index}. Consequently, the entire sky is divided into 192 independent small sky region images.

For simulated sky maps with a resolution of $N_{\rm side}=2048$, after applying the aforementioned segmentation method, each full sky map is broken down into $192$ image patches~(number of pixels at resolution $N_{\rm side}=4$), each sized $512 \times 512$ pixels, as shown in Figure~\ref{fig:pathcs}. The sky patches in the first row are used to form the training dataset, while the corresponding labels are contained within the sky patches in the second row.

At any observational scale, cosmology inherently encompasses a wealth of physical information. Assuming that we partition an full sky map into multiple independent regions where there is no scale-dependent coupling of information among these regions, any form of segmentation will inevitably entail some degree of informational loss, and our segmentation approach is no exception. Therefore, we have proposed a mitigation strategy to address this issue.

We constructed a technical framework based on a pixel-level averaging fusion method, consisting of the following key steps: First, the original full-sky CMB map is randomly rotated by predefined angles. The rotated full-sky map is then divided into map patches. Subsequently, an convolutional neural network (CNN) network architecture is applied to perform delensing on each segmented image patch, generating corresponding intermediate delensed results. After the forward processing is completed, inverse rotation operations are performed on each intermediate result to restore the original spatial orientation.

By repeating this entire process 30 times, and given that the segmentation edges vary in each iteration due to the specific rotations, the final aggregated result effectively alleviates the errors introduced by any single segmentation, thereby significantly mitigating the errors caused by the segmentation process.

This operation is applied independently to each field of the sky maps, requiring 30 rotational iterations per sky map type with corresponding repetition of the training process. The rotation angles are configured with $\phi$ values are $[-180{\rm ^\circ}, -120{\rm ^\circ},-60{\rm ^\circ}, 0{\rm ^\circ}, 60{\rm ^\circ}, 120{\rm ^\circ}]$, and $\theta$ values are $[-60{\rm ^\circ}, -30{\rm ^\circ}, -0{\rm ^\circ}, 30{\rm ^\circ}, 60{\rm ^\circ}]$.

The rotation scheme aims to mitigate the error impact caused by the segmentation operation to a certain degree, but it cannot completely eliminate it. Additionally, the implementation of multiple image rotation processes necessitates the repetition of the corresponding training process, significantly increasing the demands on computational resources and time costs. Consequently, in our research, we limited the number of rotation operations to a total of $30$ for the training dataset.

It is worth noting that since {\rm E}-mode and {\rm B}-mode represent curl-free and divergence-free polarization fields respectively, they can effectively separate signals of different physical origins. In particular, {\rm B}-mode produced by primordial gravitational waves and those induced by gravitational lensing effects can be more clearly distinguished and processed in {\rm T}/{\rm E}/{\rm B} space. Therefore, using the T/E/B representation for CMB delensing is the optimal choice. After obtaining the {\rm I}, {\rm Q}, {\rm U} Stokes parameters that have been affected by noise and instrumental effects, we can derive the {\rm T}, {\rm E}, and {\rm B} mode sky maps using Equation~(\ref{eq:t_alm}) and Equation~(\ref{eq:eb_alm}).

%%%%%%%%%%%%%%%%%%%%%%%%%%%%%%%%%%%%%%%%%%

\section{QE Delensing and UNet++ Structure}\label{sec:unet}

In the field of cosmic signal analysis and intelligent image processing, the quadratic estimator~(QE) algorithm, a secondary estimator, relies on mathematical statistics to mine deep correlations in observed signals, aiding in the inversion of cosmological parameters~\cite{Okamoto:2003zw,Schaan:2018tup,Belkner:2023duz}. The UNet++ algorithm achieves high-precision image segmentation through deep convolutional networks and multi-scale feature fusion~\cite{zhou2018unet++, zhou2019unet++}. Although the two have different paths, the former is based on physical statistical modeling, while the latter relies on data-driven learning, they jointly build a core methodology for extracting key information from complex data. We subsequently applied these two methods to delens CMB and then compared the discrepancies in their processing outcomes.

\subsection{QE delensing}

The QE delensing is a powerful technique that reconstructs and removes the gravitational‐lensing signal imprinted on the CMB by large‐scale structure, thereby sharpening acoustic features and reducing spurious B-mode power. At its core, this method uses optimal quadratic combinations of temperature ({\rm T}) and polarization ({\rm E}, {\rm B}) multipoles to estimate the lensing potential, then “undoes” the inferred deflections to recover a closer‐to‐primordial map~\cite{Diego-Palazuelos:2020lme}.

In practice, the QE algorithm reconstructs the lensing potential $\phi$ by exploiting the off-diagonal covariance induced by gravitational deflection. This deflection introduces correlations between originally independent Fourier modes of the unlensed CMB~\cite{Hu:2001kj}. Schematically, one forms weighted pairs of observed multipoles and integrates over $\ell$ to yield~\cite{Bohm:2016gzt}
\begin{equation}
	\hat{\phi}_{XY}(L) = N_L^{XY} \int \frac{{\rm d}^2l}{(2\pi)^2} X(l) Y(L - l)f(l, L) 
\end{equation}
where $X,Y\in{{\rm T},{\rm E},{\rm B}}$ and the filter $f^{XY}$ depends on theoretical $C_\ell^{\rm XY}$. The deflection field ${\rm \nabla}\phi$ mixes the unlensed E- and B-polarizations.

The normalization $N_L$ is chosen so that $\langle\phi^{{\rm TT}}\rangle= \phi$. In practice one combines all channel pairs into a minimum-variance $\hat\phi(L)$. Analogous filters $f^{XY}$ are derived for all pairs; a weighted sum of them yields the minimum-variance estimate $\hat\phi(L)$~\cite{Hu:2001kj}. 

We will perform delensing operations on all-sky temperature and polarization maps using the \href{https://github.com/NextGenCMB/delensalot/tree/main}{{delensalot}} package, and conduct a systematic quantitative analysis to compare the delensing efficacy with that of the UNet++ architecture~\cite{Belkner:2023duz}.

\subsection{UNet++ structure}

\begin{figure}
	\centering
	\includegraphics[width=0.45\textwidth]{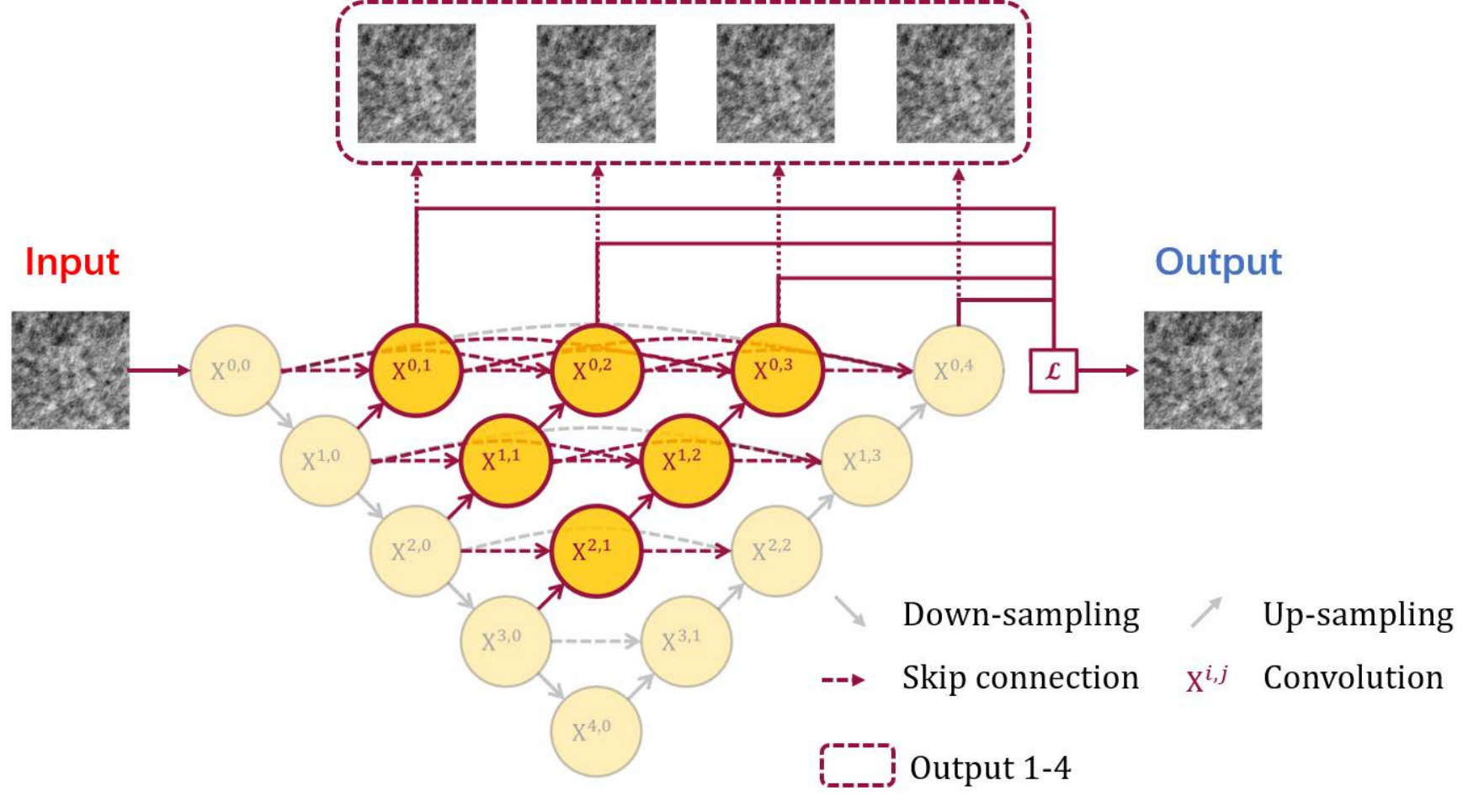}
	\caption{Unet++ network architecture. Each node in the graph represents a convolution block, downward arrows indicate down-sampling, upward arrows indicate up-sampling, dot arrows indicate skip connections, and the dot box indicates the four outputs. UNet++ combines UNets of different depths into a unified architecture. All substructures share the same encoder, but have their own decoders. Then skip connections are dropped, and every two neighboring nodes are connected with a short skip connection, enabling the deeper decoder to send supervisory signals to the shallower decoder. Finally, by connecting the decoders, a densely connected skip connection is generated so that the dense features propagate along the skip connection, resulting in more flexible feature fusion at the decoder nodes. Thus, each node in the UNet++ decoder combines multiscale features of the same resolution from all its preceding nodes from a horizontal perspective, and integrates multiscale features of different resolutions from its preceding nodes from a vertical perspective.This multiscale feature aggregation in UNet++ gradually synthesizes the segmentation, resulting in improved accuracy and fast convergence.}
	\label{fig:unet_2p}
\end{figure}

UNet++ is an enhanced version of the UNet architecture, incorporating an improved architecture to facilitate the fusion of multiscale features efficiently. The UNet network is a type of CNN that was initially designed for biomedical image segmentation~\cite{ronneberger2015u}. However, it includes considerable structural modifications based on the CNN. The primary objective is to incorporate a sequence of layers to the standard contracting network, with an upsampling operation instead of the pooling operation. As a result, the resolution of the output is increased. The expanded path is relatively symmetrical with the contracted half, generating a U-shaped structure~\cite{ronneberger2015u}.This section introduces the deep neural network architecture UNet++ \cite{zhou2018unet++, zhou2019unet++} employed in our CMB delensing analysis. In this paper, we utilize the derivative network of UNet++, as shown in Figure~\ref{fig:unet_2p}. UNet++ is an enhanced version of the UNet architecture, incorporating an improved architecture to facilitate the fusion of multiscale features efficiently. The UNet network is a type of CNN that was initially designed for biomedical image segmentation~\cite{ronneberger2015u}. However, it includes considerable structural modifications based on the CNN. The primary objective is to incorporate a sequence of layers to the standard contracting network, with an upsampling operation instead of the pooling operation. As a result, the resolution of the output is increased. The expanded path is relatively symmetrical with the contracted half, generating a U-shaped structure~\cite{ronneberger2015u}.

Specifically, the left side of the UNet is the downsampling (encoder) part, which is used to extract abstract features from the image. By using convolution and downsampling operations, the image size is reduced to extract shallow features. The convolution operation uses valid padding method, ensuring that the results are based on the context features without missing information. Therefore, after each convolution, the size of the image will be reduced.

In essence, the task of semantic segmentation entails distinguishing a particular class of images from other image classes through the utilization of segmentation masks. It can also be considered as image classification at a pixel level. Our work aims to classify the sky map of CMB temperature via convolution and convolution with the ultimate purpose of delensing CMB. As a result, our network also involves regression operations. 

Figure~\ref{fig:unet_2p} shows a unified UNet++ architecture that merges four UNets of varying depths. These UNets generate four outputs, designated output 1-4. In the graphical abstract, the original UNet appears as yellow, with skip connections depicted by dot arrows, and the four outputs displayed inside a dotted box. At the inference stage, UNet++ can be pruned by selecting a different output if it was trained with deep supervision.

By dropping some skip connections and connecting every two neighboring nodes with a short skip connection, the deeper decoder can send supervisory signals to the shallower decoder, leading to faster training. Furthermore, the decoders are connected, creating a densely connected skip connection, which allows dense features to propagate along the skip connection leading to flexible feature fusion. As a result, each node in the UNet++ decoder combines multi-scale features of the same resolution from all preceding nodes horizontally while integrating multi-scale features of different resolutions from preceding nodes vertically. This multi-scale feature aggregation in UNet++ gradually synthesizes the segmentation, resulting in improved accuracy and fast convergence.

\subsection{Loss function}

Our goal is to obtain the unlensed CMB from the lensed CMB using a supervised regression algorithm that predicts continuous output values based on input values. Therefore, we analyzed various regression loss functions, including Mean Average Error (MAE, L1 norm), Mean Squared Error (MSE, L2 norm), Huber, and Log-Cosh. Log-Cosh is a logarithmic hyperbolic cosine loss function that computes the logarithm of the hyperbolic cosine of the prediction error. When the actual value $t_i$ and the predicted value $p_i$ are given, the Log-Cosh function is defined as,
\begin{equation}\label{equ:loss_fun}
	L(p, t)=\sum_i\log\cosh(p_i-t_i).
\end{equation}

The Log-Cosh function exhibits qualities similar to those of the MAE for small losses and MSE for large losses, and features second-order differentiability. In contrast, the Huber loss function is not differentiable in all instances. MAE loss represents the average of absolute errors, and the average absolute distance between the expected and predicted data is incapable of addressing significant errors in predictions. MSE loss is the average of squared errors, and emphasizes significant errors, leading to a relatively large impact on the performance indicator. Consequently, we have chosen the log-cosh function for its superior resistance to outliers.

\subsection{Training and testing}
\begin{table}
	\centering
	\caption{Adjustment and setting of hyperparameters in the UNet++ architecture design. Prior values indicates that the optimum value is selected from the parameters of the preset value. }
	\begin{tabular}{cccc}
		\hline
		H-Param   & Description    & Prior values                     & Optimum\\
		\hline
		$\eta$            & learning rate            & [$10^{-3}, 10^{-4}, 10^{-5}$] & $10^{-4}$  \\
		$\omega$  & weight decay   & [$10^{-4}, 10^{-5}, 10^{-6}$] &$10^{-5}$   \\
		$n_{\rm filters}$ & filters  & [16, 32, 64]     & 32   \\
		$b$     & batch size & [32, 64, 128]  & 64   \\
		$\Omega$          & optimizer & [Adam, NAdam]     & NAdam   \\
		\hline
	\end{tabular}\label{tab:unet}
\end{table}

\begin{figure}
	\centering
	\includegraphics[width=0.45\textwidth]{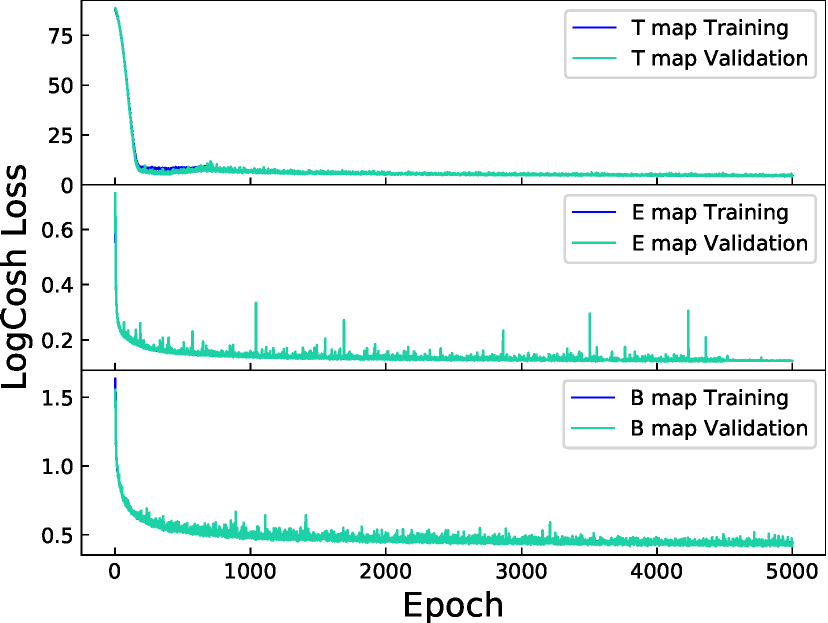}
	\caption{Loss function evolution per network over epochs. From top to bottom, represent the results of training for {\rm T}, {\rm E}, and {\rm B} maps, respectively. The dark blue solid line indicates the training set loss function evolution and the light blue solid line indicates the validation set loss function evolution.}
	\label{fig:rate_down}
\end{figure}

UNet networks are predicated on fully convolutional networks, which comprise a convolutional network and an inverse convolutional network. Thus, the heart of these networks is the convolutional layer, which involves convolving filters on the input data.

In alignment with the works of Makinen et al. and Ni et al. \cite{Makinen:2020gvh, Ni:2022kxn}, we have fixed the number of convolutional kernels at the outset of the input to $32$. The kernel size determines the convolution's field of view, which we have established at $3\times3$. To achieve the requisite output dimensionality, we have employed ``same" padding to manage sample boundaries in both convolutions and transpose convolutions. The stride parameter specifies the kernel's traversal steps across the image. In our model, we have maintained the default settings of ${\rm stride}=1$ for convolutions and ${\rm stride}=2$ for transpose convolutions.

The UNet++ architecture was used to train the CMB delensing process via a set of lensed and unlensed CMB sky maps in an end-to-end fashion.
Table~\ref{tab:unet} displays the hyperparameters' specifics used in this network.
The NAdam optimizer was utilized in the analysis with the standard TensorFlow parameters~\cite{ruder2016overview, Sashank:2019abs}.

The hyperparameters were meticulously fine-tuned for network optimization.
The batch size and the initial number of convolutional filters are optimized to $32$, $64$ and $128$, respectively, being restricted by GPU memory.
The number of epochs is fixed to $3000$; Table~\ref{tab:unet} illustrates the learning rate setup.
Meanwhile, weight decay was examined with a list of previous values listed in Table~\ref{tab:unet}.
The optimized values of the initial number of convolution filters, learning rate, weight decay, and batch size were fixed at $32$, $10^{-4}$, $10^{-5}$, $64$, respectively. The entire number of trainable parameters is $7.4\times10^{7}$. 
The total number of trainable parameters is $9.04\times10^{6}$. 
Since the CMB sky map data contains both positive and negative values due to temperature fluctuations, ReLU and LeakyReLU activation with alpha=1.0 was chosen to handle this relationship.
Figure~\ref{fig:rate_down} illustrates the evolution of the loss function during the training of {\rm T}, {\rm E}, and {\rm B} maps as a function of the number of epochs. The figure shows the training results of {\rm T}, {\rm E}, and {\rm B} maps in a top-to-bottom order. The dark blue line indicates the evolution of the training set loss function, while the light blue dashed line represents the validation dataset's loss function development.

In this study, we utilized a high-performance computing platform equipped with eight NVIDIA A40 GPUs, an Intel Xeon Platinum 8358 processor, and 1000 GB of memory, providing robust computational support for the training of complex models. Taking a set of parameter-generated sky maps (e.g., the {\rm T} field) as an example, a single training run takes approximately 16 hours. For datasets containing all three fields ({\rm T}, {\rm E}, and {\rm B}), after applying 30 different rotational transformations, a total of 90 training tasks were performed, accumulating to 1440 hours of total training time.

%%%%%%%%%%%%%%%%%%%%%%%%%%%%%%%%%%%%%%%%%%

\section{Results and Discussion}\label{sec:result}

In this section, we conducted a thorough analysis of the results generated by the QE algorithm and the UNet++ model. Specifically, we systematically explored the outcomes of these two methods from two key perspectives: sky map patches and power spectra.
\subsection{Sky map analysis}
\begin{figure*}
	\centering
	\includegraphics[width=0.8\textwidth]{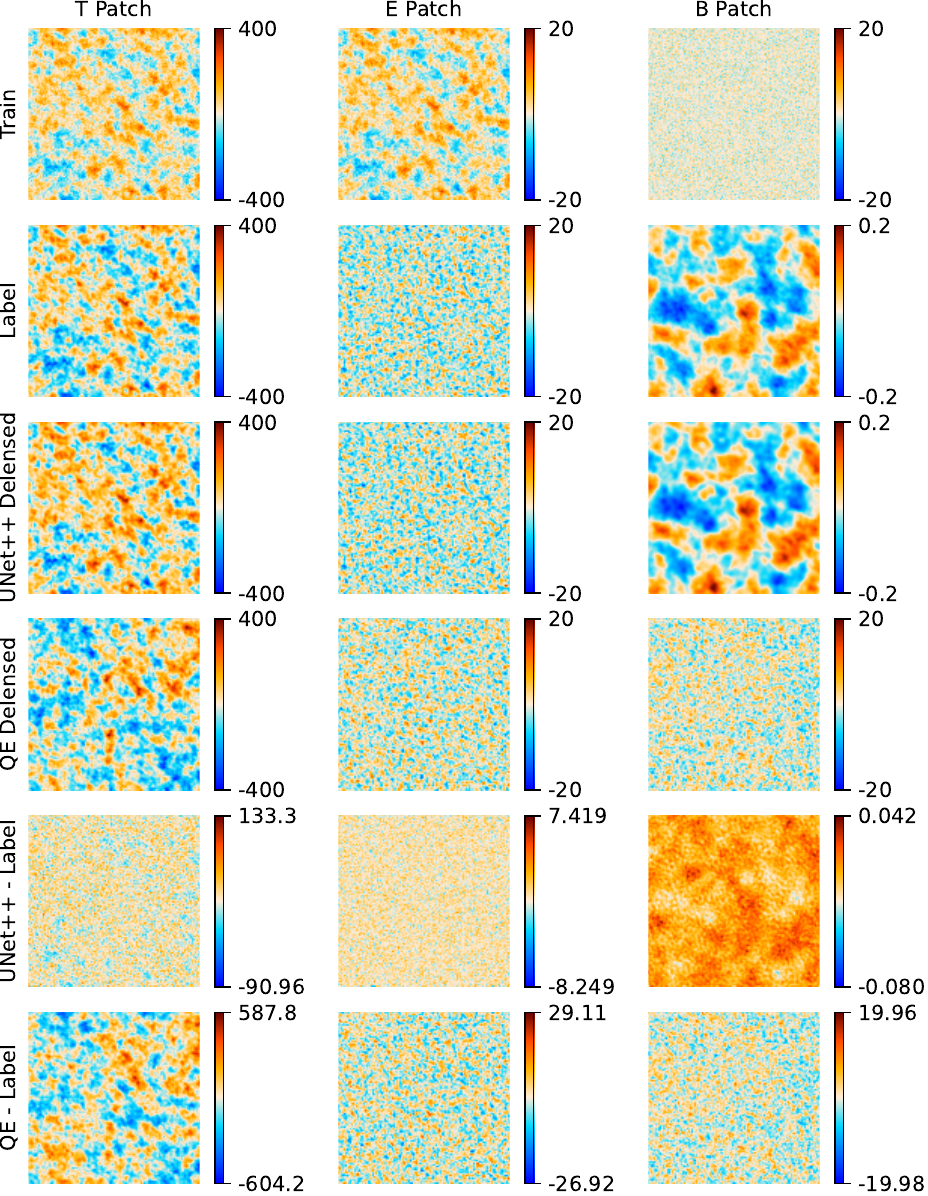}
	\caption{Comparison of training and predicted sky map patches. The image is arranged in columns corresponding to the three sky map patches {\rm T}, {\rm E}, and {\rm B} in order. The columns are arranged from top to bottom, with the first row shows the sky map patch that has been lensed and is affected by noise and instrumental effects. the second row showing the original sky patch of the real sky image that is not affected by gravitational lensing effects, the third row showing the lens-removed prediction result of the UNet++ model after processing, the fourth row showing the lens-removed prediction result of the QE algorithm after processing, the fifth row showing the residual map of the UNet++ prediction result and the real image, and the sixth row showing the residual map of the QE algorithm prediction result and the real image. Similar to Figure~\ref{fig:pathcs}, the size of a sky patch is $214.86~{\rm deg}^2$, with the center located at $(l,b)=(78.75{\rm ^\circ}, 0{\rm ^\circ})$. The unit is ${\rm \mu K}$.}
	\label{fig:pathcs_pred}
\end{figure*}

Based on the network parameter configuration described in the previous chapter, we have trained multiple datasets and performed prediction operations on the test set. Figure~\ref{fig:pathcs_pred} shows the residual performance between the predicted and true values of UNet++ model processing results. 
Firstly, the first and second rows of the modified figure display a sky map patch from the test set, which includes lensing effects, noise, and instrumental effects, alongside its corresponding label (an unlensed sky map patch). Secondly, the third and fourth rows present the delensing effect visualizations generated by the UNet++ and QE algorithms, respectively. Finally, the fifth and sixth rows illustrate the differences between the delensing results produced by the UNet++ and QE algorithms and their respective labels.

It can be clearly observed from the analysis of Figure~\ref{fig:pathcs_pred} that the delensing method we used has played a significant role in eliminating the lensing effect. To quantitatively evaluate the results, we used two image quality evaluation metrics, namely structural similarity (SSIM) and peak signal-to-noise ratio (PSNR).

SSIM is mainly used to analyze the structural information of two images~\cite{wang2004image,wang2009mean}. The closer the SSIM is to $1$, the more similar the structural information of the two images is. We have generally processed SSIM by setting the parameters $\alpha, \beta, \gamma$ to $1$. For the sky patches of two fields $\mathcal{F}_1$ and $\mathcal{F}_2$, we have carried out the following treatment:
\begin{equation}\label{eq:ssim}
	\begin{split}
		{\rm SSIM}(\mathcal{F}_1, \mathcal{F}_2) &= [l(\mathcal{F}_1, \mathcal{F}_2)^\alpha \cdot c(\mathcal{F}_1, \mathcal{F}_2)^\beta \cdot s(\mathcal{F}_1, \mathcal{F}_2)^\gamma]\\
		&= \frac{(2\bar{\mathcal{F}}_1\bar{\mathcal{F}}_2+c_1)[2{\rm CoV}({\mathcal{F}_1, \mathcal{F}_2})+c_2]}{(\bar{\mathcal{F}}_1\bar{\mathcal{F}}_2^2+c_1)[\sigma(\mathcal{F}_1)^2+\sigma(\mathcal{F}_1)^2+c_2]},
	\end{split}
\end{equation}
where $l(\mathcal{F}_1, \mathcal{F}_2)$, $c(\mathcal{F}_1, \mathcal{F}_2)$, $s(\mathcal{F}_1, \mathcal{F}_2)$ represent the brightness comparison, contrast comparison, and structure comparison between $\mathcal{F}_1$
and $\mathcal{F}_2$, respectively. ${\rm CoV}(\mathcal{F}_1, \mathcal{F}_2)$ represents the covariance between $\mathcal{F}_1$ and $\mathcal{F}_2$, while $\sigma(\mathcal{F}_1)$ and $\sigma(\mathcal{F}_2)$ represent the standard deviation of $\mathcal{F}_1$ and $\mathcal{F}_2$, respectively. $c1$ and $c_2$ are constants to prevent the numerator or denominator from being zero. 

According to Equation~(\ref{eq:ssim}), we can give the value of SSIM, as shown in Table~~\ref{tab:ssim}. From the table, it can be seen that the SSIM values of the {\rm T}, {\rm E}, and {\rm B} sky map patches after delensing processing using the UNet++ algorithm are almost close to $1$. This means that the image blocks of the sky map after lens processing using the UNet++ algorithm are very similar in structure to those without lens.

In contrast, the results of the QE algorithm are much lower than those obtained by the UNet++ algorithm. Especially in the power spectrum of B-mode, this difference is more obvious. This can be explained by the fact that the UNet++ algorithm is better than the QE algorithm in removing lensing effects.

\begin{table}
	\centering\caption{SSIM values between predicted map patches and ground truth labels.}\label{tab:ssim}
	\begin{tabular}{|c|c|c|c|}  
		\hline 
		~& ${\rm SSIM}_{({\rm T}, {\rm T})}$ & ${\rm SSIM}_{({\rm E}, {\rm E})}$ & ${\rm SSIM}_{({\rm B}, {\rm B})}$ \\ \hline 
		${\rm Patch}_{({\rm Truth}, QE)}$ & 0.7667 & 0.7390 & 0.0282 \\ \hline 
		${\rm Patch}_{({\rm Truth}, UNet++)}$ & 0.9880 & 0.9760 & 0.9544 \\ \hline 
	\end{tabular}  
\end{table}

As an image quality assessment metric, SSIM fully considers the sensitivity of the human eye to structural information and focuses on differences in image structural information, thus being able to more accurately reflect the quality of the image. However, when comprehensively evaluating image quality, we still need to pay attention to pixel-level errors. Therefore, we further calculate the PSNR value between different images, using the following specific formula:
\begin{equation}
	PSNR(\mathcal{F}_1 - \mathcal{F}_2) = 10 \cdot\log_{10}[\frac{{\rm MAX}(\mathcal{F}_1, \mathcal{F}_2)}{{\rm MSE}(\mathcal{F}_1, \mathcal{F}_2)}]{\rm dB},
\end{equation}
where ${\rm MAX}(\mathcal{F}_1, \mathcal{F}_2)$ represents the maximum pixel difference between images $\mathcal{F}_1$ and $ \mathcal{F}_2$, and ${\rm MSE}(\mathcal{F}_1, \mathcal{F}_2)$ represents the mean squared error between $\mathcal{F}_1$ and $ \mathcal{F}_2$. Through this calculation, we can more comprehensively evaluate the degree of distortion at the pixel level in the image, and thus provide strong support for the comprehensive evaluation of image quality. 

PSNR stands as a simple yet widely utilized metric for assessing image quality, employing the decibel (dB) as its unit of measurement~\cite{jahne2005digital,bovik2010handbook}. Within this evaluative framework, a higher numerical value signifies a lower degree of image distortion, thereby indicating superior image quality. Akin to the computation of the SSIM metric, we have also conducted a corresponding analytical calculation of the PSNR, as shown in Table~\ref{tab:psnr}. 
\begin{table}
	\centering\caption{PSNR values between predicted map patches and ground truth labels. The unit is ${\rm dB}$.}\label{tab:psnr}
	\begin{tabular}{|c|c|c|c|}  
		\hline 
		~ & ${\rm PSNR}_{({\rm T}, {\rm T})}$ & ${\rm PSNR}_{({\rm E}, {\rm E})}$ & ${\rm PSNR}_{({\rm B}, {\rm B})}$ \\ \hline 
		${\rm Patch}_{({\rm Truth}, QE)}$ & 21.14 & 19.25 & 14.48 \\ \hline 
		${\rm Patch}_{({\rm Truth}, UNet++)}$ & 37.70 & 38.76 & 37.87 \\ \hline 
	\end{tabular}  
\end{table}

By analyzing the data in Table~\ref{tab:psnr}, we can find that the PSNR ratio obtained by the UNet++ algorithm is also significantly higher than that of the QE algorithm. This result shows that the UNet++ algorithm performs better in signal recovery and can restore the true signal more effectively.

\subsection{Angular power spectrum analysis}

In the previous data simulation stage, in order to study the impact of the lensing effect on the CMB, we conducted 30 simulations for each component. Considering that the segmentation operation of the sky map may introduce errors in scale information, in order to reduce the errors caused by this segmentation, we performed a specific angle rotation on the simulated sky map, followed by a delensing operation, and finally rotated back to the original angle. After such processing, we averaged the results to obtain a complete sky map for subsequent angular power spectrum calculations.

\begin{figure}
	\centering
	\includegraphics[width=0.45\textwidth]{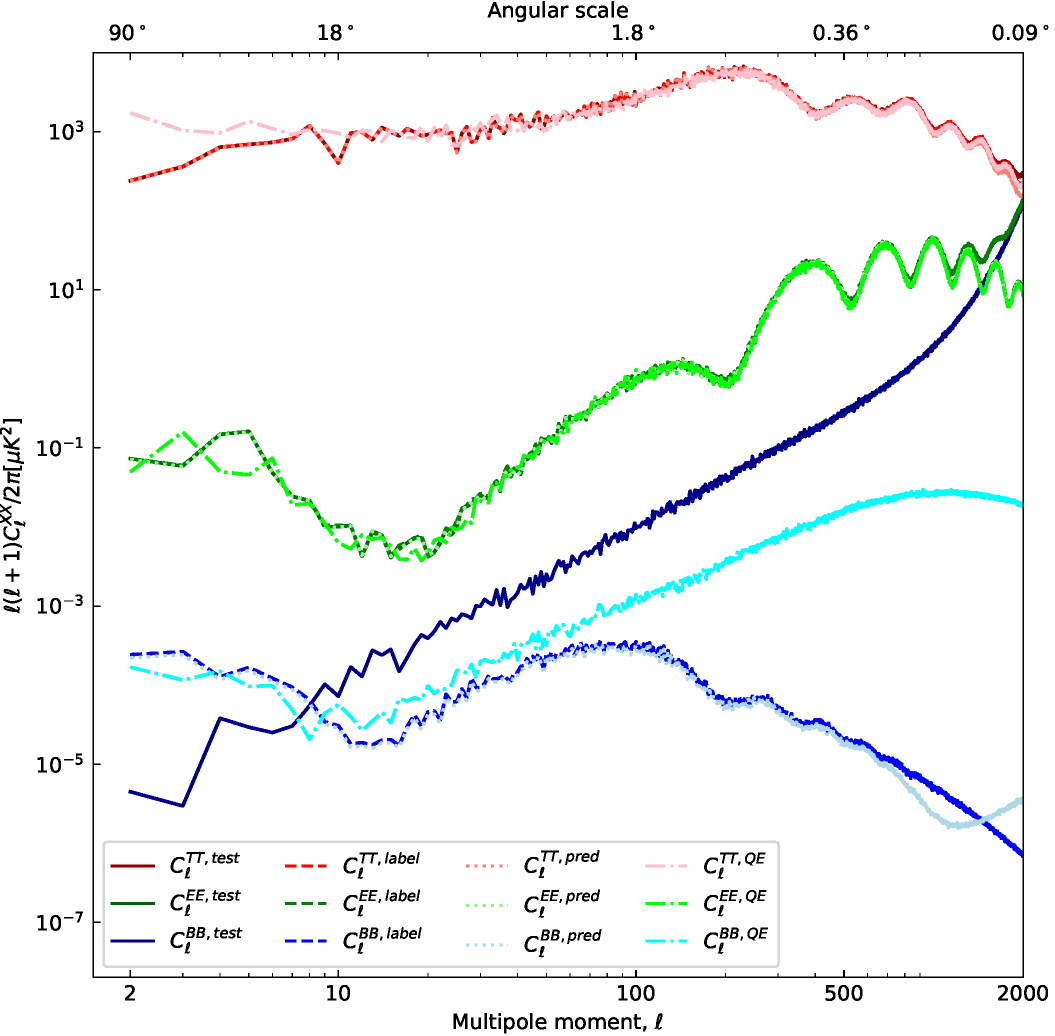}
	\caption{For the {\rm TT} spectrum, the dark red solid line denotes the lensed power spectrum, the red dashed line shows the unlensed spectrum, the salmon dash-dot line corresponds to the spectrum delensed using UNet++, and the pink dotted line represents delensing via the QE algorithm.
		In the {\rm EE} panel, the dark green solid line indicates the lensed {\rm EE} power spectrum, the green dashed line shows the unlensed case, the light green dash-dot line represents delensing with UNet++, and the lime dotted line shows the delensed spectrum obtained by the QE method.
		For the {\rm BB} spectrum, the dark blue solid line denotes the lensed {\rm BB} power spectrum, the blue dashed line represents the unlensed primordial signal, the slight blue dash-dot line corresponds to delensing by UNet++, and the cyan dotted line shows the result from QE-based delensing. The parameters for the temperature power spectrum are set to $A_s = 2.1\times10^{-9}$ and $r = 0.005$.}
	\label{fig:result_single_TEB}
\end{figure}

To more intuitively present the impact of rotation processing on the results, we first show the power spectra of each component throughout the day without rotation processing, as shown in Figure~\ref{fig:result_single_TEB}. In this figure, we present the {\rm TT}, {\rm EE}, and {\rm BB} power spectra of the CMB without lensing effect processing. The temperature power spectrum is represented by a dark red solid line, which represents the angular power spectrum generated by the {\rm T} signal without lens processing. The long red dashed line represents the angular power spectrum generated by the {\rm T} signal without lens processing through the UNet++ network. The light red dotted line represents the angular power spectrum generated by the {\rm T} signal without lens processing through the QE algorithm. The polarization power spectrum covers the {\rm EE} and {\rm BB} spectra, and is represented by dark blue, blue long dashed lines, light blue dotted lines, dark green, green long dashed lines, and light green dotted lines in a similar manner to represent the angular power spectra generated by the {\rm T} signal without lens processing, UNet++ lens processing, and QE algorithm lens processing.

As can be seen from Figure~\ref{fig:result_single_TEB}, there are significant differences between the delensing effects of the UNet++ and QE algorithms at high $\ell$ values (small scales) and the label power spectrum. This phenomenon reveals the varying performance of the QE and UNet++ network across different scales. Specifically, as the scale decreases, the delensing effect of UNet++ diminishes gradually, and the rate of this decrease accelerates. We propose two possible explanations for this finding.

Firstly, we believe that the lensing effect may gradually increase on smaller scales, causing a significant distortion of the original CMB signal. This distortion may exceed the correction capability of the UNet++ network, leading to a decrease in delensing effectiveness on even smaller scales.
Secondly, after dividing the full-sky map into small patches and performing delensing processing on them separately, the algorithms can reconstruct the central regions of each patch relatively accurately, but often fail to fully recover or show slight misalignments at the edges. After stitching the patches together, the tiny discontinuities at the boundaries of each patch correspond to the high $\ell$ components in the angular power spectrum. In contrast, low $\ell$ modes span multiple small patches, and the algorithms tend to retain the average values of each patch. Therefore, even if there are subtle offsets in the overall large - scale structure after stitching, it can still maintain good continuity. To reduce the impact of boundary artifacts on high $\ell$ signals, we attempt to perform rotation processing on the small patches.

It is worth noting that the two algorithms also exhibit opposite deviations in the high $\ell$ range. This is because UNet++, in order to suppress noise amplification, applies stronger smoothing or regularization to strong signals, resulting in a power spectrum recovered at high $\ell$ values that is lower than the label. On the other hand, the QE algorithm is more inclined to preserve all the details, leading to a power spectrum at high $\ell$ values that is higher than the label.

\begin{figure}
	\centering
	\includegraphics[width=0.45\textwidth]{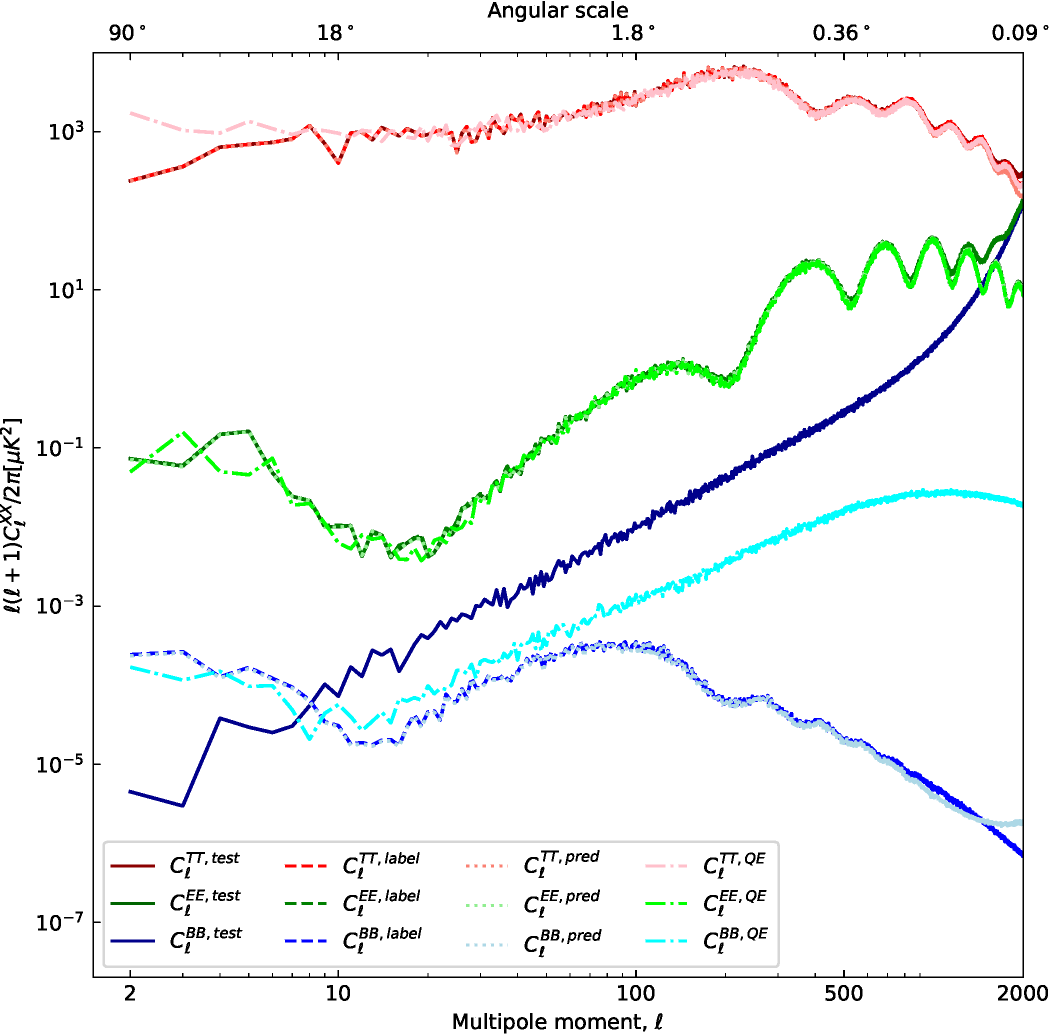}
	\caption{Angular power spectrum of the unlensed and predicted CMB {\rm TT}, {\rm EE} and {\rm BB}. The results of UNet++ are obtained by averaging over 30 rotational transformations, while the power spectra of the QE algorithm are calculated based on the full-sky map. The line color and line style in this diagram are consistent with Figure~\ref{fig:result_single_TEB}.}
	\label{fig:result_mutil_TEB}
\end{figure}

To optimize the delensing results,as previously stated, we initially applied 30 rotational transformations to the acquired full-sky maps. Subsequently, we mitigated the lensing effects present in the rotated data. Finally, we applied another rotational operation to the processed data, culminating in the results depicted in Figure~\ref{fig:result_mutil_TEB}. Among them, the results obtained through the QE delensing algorithm are directly processed based on the full-sky map. Consequently, no additional rotation and segmentation operations are required for this result. As evident from the figure, the errors stemming from image segmentation have been substantially reduced following the rotational processing.

In order to more intuitively compare the effects of delensing on various scales, we introduce two quantities to quantify the results of this process, which are specifically defined as follows:

\begin{equation}\label{eq:index}
	\begin{split}
		\Gamma^{XX}_{\Lambda}={\rm Mean}(|\frac{C_\ell^{XX_{unlensed}}}{C_\ell^{XX_{delensed}}}-1|)
	\end{split}
\end{equation}
where the lower subscript $\Lambda$ represents the QE or the UNet++ algorithm, and the upper subscript $XX=\{{\rm TT}, {\rm EE}, {\rm BB}\}$. The closer the value of $\Gamma$ is to zero, the better the delensing effect; Conversely, the greater the deviation of $\Gamma$ from zero, the less satisfactory the delensing effect.

Based on the quantitative evaluation metric specified in Equation~(\ref{eq:index}), for the {\rm TT} spectrum, the error value of the QE algorithm is $\Gamma^{TT}_QE=0.6451$, while that of the UNet++ algorithm is $\Gamma^{TT}_UNet++=0.0582$. This indicates that the error of the QE algorithm is approximately 10 times that of the UNet++ algorithm. For the {\rm EE} spectrum, the error of the QE algorithm is $\Gamma^{EE}_QE=0.4560$, and the error of the UNet++ algorithm is $\Gamma^{EE}_UNet++=0.0362$. Here, the error of the QE algorithm is roughly 11 times that of the UNet++ algorithm. In the case of the {\rm BB} spectrum, the disparity in errors between the two algorithms is even more pronounced. The error of the QE algorithm is $\Gamma^{BB}_QE=0.9224$, and the error of the UNet++ algorithm is $\Gamma^{BB}_UNet++=0.0693$. At this point, the error of the QE algorithm is approximately 12 times that of the UNet++ algorithm. Based on these comparisons, it is evident that the performance of UNet++ exhibits a significant improvement over the QE algorithm.

\subsection{Discussion}

The results and analysis presented in the previous section demonstrate that the UNet framework achieves remarkably high performance in delensing. This outcome arises from the deliberate design of our proof-of-concept experiment: the network was trained to map noisy, lensed inputs onto idealized B-mode maps that are both noiseless and unlensed. Consequently, the model is encouraged not only to remove distortions induced by gravitational lensing but also to suppress noise, leading to results that appear exceptionally strong.

A further concern is whether such performance merely reflects memorization of the training labels rather than the learning of physically meaningful patterns. To address this, we enforced strict separation between training, validation, and test datasets, and ensured that the cosmological parameters used to generate the test maps differed from those employed in training. The results indicate that the network performs well when the test parameters lie close to the training distribution, whereas its performance degrades for out-of-distribution cases. This distribution-dependent behavior suggests that the model acts analogously to an interpolation scheme: it achieves strong results within the range of the training data but exhibits limited extrapolation capability. Crucially, this behavior confirms that the network has not simply memorized the training labels, but has instead learned nontrivial mappings associated with the gravitational lensing process, effectively ruling out overfitting.

It should be emphasized that our current experiments are entirely based on simulated data. In the process of simulation, certain complexities inevitably present in real observations are omitted, such as non-Gaussian noise, residual foreground contamination, instrumental systematics, and uncertainties related to cosmological models. Therefore, we do not rule out the possibility that the model’s performance may degrade when applied to real observational data. Nevertheless, delensing of the CMB remains a frontier research area. In the absence of large-scale, high-quality observational data, exploring and validating new methods using simulated data is an essential step toward advancing this field.

Looking forward, with the upcoming high-quality observational data from next-generation experiments such as {\it LiteBIRD}, {CMB-S4}, {AliCPT},
and {PIPER}, it will become possible to validate and refine these deep learning approaches under conditions closer to real observations. Furthermore, by incorporating more accurate noise modeling, complex foreground components, and instrumental systematics, future studies are expected to systematically evaluate and potentially extend the applicability and robustness of the methods proposed in this work.

%%%%%%%%%%%%%%%%%%%%%%%%%%%%%%%%%%%%%%%%%%
\section{Conclusions}\label{sec:conclusions}

In CMB observations, weak gravitational lensing distorts CMB images, causing mixing between the T-mode and E-mode, as well as between the E-mode and B-mode. Therefore, finding effective methods to remove the effects of weak gravitational lensing on CMB data is one of the key challenges in revealing the true B-mode polarization signal.

We have adopted a relatively novel solution that allows spherical data to be processed using deep learning methods. 

Our approach presents several advantages. It can be applied regardless of the resolution of the full-sky map and requires no complex initialization procedures. The data can be used for model training after simple preprocessing of the sky maps. Segmentation errors are inevitably present in sky maps irrespective of the data type but our model retains the original data resolution as much as possible through multiple rotation operations.

During the delensing study of full-sky images of CMB T-mode, polarization E-mode, and B-mode components, the UNet++ network demonstrated outstanding performance. We analyzed it at both the image and angular power spectrum levels.

At the image level, we employed two image quality assessment metrics, SSIM and PSNR. For map patches of {\rm T}/{\rm E}/{\rm B} modes, the UNet++ algorithm showed values of SSIM very close to 1 compared with the QE algorithm, and PSNR values were also within a highly similar range (greater than $30$ dB).

At the power spectrum level, we analyzed the characteristic changes in {\rm TT}, {\rm EE}, and {\rm BB} spectra. By calculating corresponding quantitative indicators, the value of $\Gamma^{TT}_UNet++$ was only 0.0582, significantly lower than the delensed $\Gamma^{TT}_QE = 0.6451$, which fully demonstrates the significant effectiveness of the delensing method. Similarly, there was a marked improvement observed in the {\rm EE} spectrum: $\Gamma^{EE}_UNet++ = 0.0362$, while the corresponding $\Gamma^{EE}_QE = 0.4560$. The difference between lensed and unlensed {\rm BB} spectra was particularly pronounced, hence its delensing effect was even more prominent. Specifically, $\Gamma^{BB}_UNet++ = 0.0693$, whereas $\Gamma^{BB}_QE$ approached $1$, further confirming the superior performance of the UNet++ algorithm in this mode. Moreover, by introducing a rotation mechanism, the angular power spectrum improved significantly compared to the non-rotated scenario. However, there is still room for optimizing the delensing effect on small-scale structures.

Furthermore, weak cosmological signals, especially B-mode polarization signals, are particularly susceptible to contamination by foreground radiation. Therefore, in future work, we will consider the impact of foreground components to achieve effective subtraction of CMB foregrounds.

\section*{Data Availability}

The code used to perform the analysis in this work is publicly available at \url{https://github.com/understars0516/cmb_delensing}. The repository includes all necessary scripts, configuration files, and instructions to reproduce the key results and figures presented in this paper.

\section*{Acknowledgments}
We thank Sebastian Belkner for his valuable suggestions and guidance on the correct use of Lenspyx. This work was supported by the National SKA Program of China (Grants Nos. 2022SKA0110200 and 2022SKA0110203), the National Natural Science Foundation of China (Grants Nos. 12533001, 12575049, 12473001, 11975072, 11835009, and 11875102), the Liaoning Revitalization Talents Program (Grant No. XLYC1905011), the National 111 Project of China (Grant No. B16009), and the Science Research Grants from the China Manned Space Project (Grant No. CMS-CSST-2025-A02).

\bibliography{main}% Produces the bibliography via BibTeX.

\end{document}